\documentclass{aa}
\usepackage{graphicx}
\usepackage{txfonts}
\usepackage{hyperref}
\usepackage{float}
\hypersetup{
    colorlinks,
    citecolor=blue,
    filecolor=blue,
    linkcolor=blue,
    urlcolor=blue}

\newcommand{\sg}{S_{\rm G}}
\newcommand{\sgone}{S_{\rm G1}}
\newcommand{\sgtwo}{S_{\rm G2}}
\newcommand{\sn}{S_{\rm N}}
\newcommand{\snone}{S_{\rm N1}}
\newcommand{\sntwo}{S_{\rm N2}}
\newcommand{\sa}{S_{\rm A}}

\newcommand{\fra}{F_{\rm 10.7}}
\newcommand{\fca}{F_{\rm CaIIK}}
\newcommand{\fla}{F_{L\alpha}}
\newcommand{\fmg}{F_{\rm MgII}}
\newcommand{\fmf}{F_{\phi}}
\newcommand{\mfs}{F^{*}\left(S\right)}
\newcommand{\mfstvm}{F^{*}_{\rm TVM}\left(S\right)}
\newcommand{\mfspw}{F^{*}_{\rm PW}\left(S\right)}
\newcommand{\mfspwemp}{F^{*}_{\rm PW}\left(\sa\right)}
\newcommand{\mfsysk}{F^{*}_{\rm YSK}\left(S\right)}

\newcommand{\mfsaysk}{F^{*}_{\rm YSK}\left(\sa\right)}
\newcommand{\mfsnoneysk}{F^{*}_{\rm YSK}\left(\snone\right)}
\newcommand{\mfsntwoysk}{F^{*}_{\rm YSK}\left(\sntwo\right)}
\newcommand{\mfsgoneysk}{F^{*}_{\rm YSK}\left(\sgone\right)}
\newcommand{\mfsgtwoysk}{F^{*}_{\rm YSK}\left(\sgtwo\right)}
\newcommand{\mfcasaysk}{F^{*}_{\rm CaIIK,YSK}\left(\sa\right)}
\newcommand{\mflasaysk}{F^{*}_{L\alpha{\rm ,YSK}}\left(\sa\right)}
\newcommand{\mfmgsaysk}{F^{*}_{\rm MgII,YSK}\left(\sa\right)}

\newcommand{\hpwemp}{H_{\rm PW,emp}}
\newcommand{\hpwmod}{H_{\rm PW,mod}}
\newcommand{\hysk}{H_{\rm YSK}}
\newcommand{\fitone}{\sigma^2_{\bot}/\sigma^2_{\parallel}}
\newcommand{\fittwo}{1-R^2}
\newcommand{\fitthree}{{\rm RMS}_{\rm NR}}
\newcommand{\smoothf}{\left<F\right>_{\rm 3Y}}
\newcommand{\smoothfmax}{\left<F\right>_{\rm 3Y,2000}}
\newcommand{\smoothfmin}{\left<F\right>_{\rm 3Y,2008}}
\newcommand{\smoothmfs} {\left<\mfs\right>_{\rm 3Y}}
\newcommand{\fb}{{\rm FB}}
\newcommand{\sd}{{\rm SD}}
\newcommand{\tsi}{{\rm TSI}}
\newcommand{\ps}{{\rm PSI}}
\newcommand{\af}{A\left(F\right)}
\newcommand{\as}{A\left(S\right)}
\newcommand{\asa}{A\left(\sa\right)}
\newcommand{\asnone}{A\left(\snone\right)}
\newcommand{\amfsysk}{A\left(\mfsysk\right)}
\newcommand{\amfsaysk}{A\left(\mfsaysk\right)}
\newcommand{\amfsnoneysk}{A\left(\mfsnoneysk\right)}
\newcommand{\afb}{A\left(\fb\right)}
\newcommand{\asd}{A\left(\sd\right)}
\newcommand{\atsi}{A\left(\tsi\right)}
\newcommand{\ratio}{\amfsysk/\as}
\newcommand{\ratiosa}{\amfsaysk/\asa}
\newcommand{\ratiosnone}{\amfsnoneysk/\asnone}
\newcommand{\ratiofbsd}{\afb/\left|\asd\right|,}
\newcommand{\logrhk}{\log{}R'_{\rm HK}}

\begin{document}

\title{How faculae and network relate to sunspots, and the implications for solar and stellar brightness variations}
\titlerunning{How faculae and network relate to sunspots}
\author{K. L. Yeo\inst{1} \and S.~K.~Solanki\inst{1,2} \and N.~A.~Krivova\inst{1}}
\institute{Max-Planck Institut f\"{u}r Sonnensystemforschung, Justus-von-Liebig-Weg 3, 37077 G\"{o}ttingen, Germany\\ \email{yeo@mps.mpg.de} \and School of Space Research, Kyung Hee University, Yongin, 446-701 Gyeonggi, Korea}
\abstract
{How global faculae and network coverage relates to that of sunspots is relevant to the brightness variations of the Sun and Sun-like stars.}
{We aim to extend and improve on earlier studies that established that the {facular-to-sunspot-area} ratio diminishes with total sunspot coverage.}
{Chromospheric indices and the total magnetic flux enclosed in network and faculae, referred to here as `facular indices', are modulated by the amount of facular and network {present}. We probed the relationship between various facular and sunspot indices through an empirical model, taking into account how active regions evolve and the possible non-linear relationship between plage emission, facular magnetic flux, and sunspot area. This model was incorporated into a model of total solar irradiance (TSI) to elucidate the implications for solar and stellar brightness variations.}
{The reconstruction of the facular indices from the sunspot indices with the model {presented} here replicates most of the observed variability, and is better at doing so than earlier models. Contrary to recent studies, we found the relationship between the facular and sunspot indices to be stable over the past four decades. The model indicates that, like the {facular-to-sunspot-area} ratio, the ratio of the variation in chromospheric emission and total network and facular magnetic flux to sunspot area decreases with the latter. {The TSI model indicates the ratio of the TSI excess from faculae and network to the deficit from sunspots also declines with sunspot area,} with the consequence being that TSI rises with sunspot area more slowly than if the two quantities were linearly proportional to one another. This explains why even though solar cycle 23 is significantly weaker than cycle 22, TSI rose to comparable levels over both cycles. The extrapolation of the TSI model to higher activity levels indicates that in the activity range where Sun-like stars are observed to switch from growing brighter with increasing activity to becoming dimmer instead, the activity-dependence of TSI exhibits a similar transition. This happens as sunspot darkening starts to rise more rapidly with activity than facular and network brightening. This bolsters the interpretation of this behaviour of Sun-like stars as the transition from a faculae-dominated to a spot-dominated regime.}
{}
\keywords{Sun: activity - Sun: faculae, plages - Sun: magnetic fields - sunspots}
\maketitle

\section{Introduction}
\label{introduction}

The variation in solar irradiance at timescales greater than a day is believed to be dominantly driven by photospheric magnetism \citep{solanki13,yeo17b}. Models developed to reproduce solar irradiance variability by relating it to magnetic activity on the solar surface provide the radiative forcing input required by climate simulations \citep{haigh07}. Solar irradiance variability is modelled as the sum effect of the intensity deficit from sunspots and the excess from faculae and {network}, determined from observations of solar magnetism \citep{domingo09,yeo14b}. Most of the models aimed at reconstructing solar irradiance variability back to pre-industrial times, a period of particular interest to climate studies, rely on sunspot indices such as the total sunspot area, international sunspot number, and group sunspot number, as these are the only direct observations of {solar magnetic} features to go this far back in time \citep[e.g.][]{lean00,krivova07,krivova10,dasiespuig14,dasiespuig16,coddington16,wu18}. Of course, inferring the effect of not just sunspots but also of faculae and {network} on solar irradiance from sunspot indices requires knowledge of how the {amount} of faculae and {network} present relates to sunspots. For example, the solar irradiance reconstruction by \cite{dasiespuig14,dasiespuig16} made use of the model by \cite{cameron10}, which incorporates the empirical relationship between facular and sunspot area reported by \cite{chapman97}, to calculate the {amount} of faculae and {network} present from the sunspot area and number.

The understanding of how faculae and {network} relate to sunspots is also relevant to that of the brightness variations of Sun-like stars. The synoptic programmes at the Fairborn, Lowell, and Mount Wilson observatories monitored the brightness and activity of a number of Sun-like stars as indicated by the Str{\"o}mgren $b$ and $y$ (i.e. visible) photometry and Ca II H\&K emission (as a proxy of activity). These observations revealed a dichotomy in the relationship between brightness and activity. While the brightness of less active, older stars rises with increasing activity, it diminishes for more active, younger stars \citep{lockwood07,hall09,shapiro14,radick18}. This switch in activity-dependence is interpreted as the transition from a faculae-dominated regime, where the intensity excess from faculae and {network} has a greater effect on brightness variations than the deficit from starspots, to a spot-dominated regime where the converse is true. The threshold between the two regimes is estimated to be at $\logrhk$ \citep[see definition in][]{noyes84} of between $-4.9$ and $-4.7$ \citep{lockwood07,hall09,shapiro14,radick18}. The Sun has a mean $\logrhk$ of about $-4.9$ \citep[see][and Sect. \ref{analysis3}]{lockwood07} and appears to be faculae-dominated \citep{shapiro16,radick18}, suggesting that it lies not far below this threshold. The transition between the faculae-dominated and spot-dominated regimes can therefore be probed by looking at how solar faculae and {network} relate to sunspots and by extrapolating the apparent relationship to higher activity levels.

\cite{chapman97} examined the relationship between total sunspot area and Ca II K plage area, taken as a proxy of facular area, over the declining phase of solar cycle 22. This makes use of the fact that chromospheric emission is strongly enhanced in plage and network features overlaying photospheric faculae and {network} \citep[e.g.][]{schrijver89,harvey99,loukitcheva09,kahil17,barczynski18}. \cite{chapman97} found that facular area conforms to a quadratic relationship with sunspot area. The coefficient of the second-order term is negative, such that the facular-to-sunspot-area ratio decreases as sunspot area increases. Investigations by \cite{foukal93,foukal96,foukal98} and \cite{shapiro14}, extending multiple solar cycles, returned similar results. However, while \cite{chapman97} made use of modern, relatively pristine Ca II K spectroheliograms, \cite{foukal93,foukal96,foukal98} looked at facular areas based on historical Ca II K spectroheliograms, which suffer from calibration issues and defects \citep{ermolli09}, and white-light heliograms, where the intensity contrast of faculae is not only weak, but {also} diminishes towards the disc centre \citep{foukal04}. While \cite{chapman97} and \cite{foukal93,foukal96,foukal98} made use of measured facular area, \cite{shapiro14} examined the total {facular} and network disc coverage from \cite{ball12}, which was determined indirectly from full-disc magnetograms using an empirical relationship between the magnetogram signal and the facular filling factor \citep{fligge00}. A proper examination of the relationship between sunspot and facular area over multiple solar cycles is still lacking. The restoration and calibration of the various historical Ca II K spectroheliogram archives, which extend as far back as the beginning of the last century, will facilitate such investigations \citep{chatzistergos18,chatzistergos19a,chatzistergos19b}.

As noted in the previous paragraph, chromospheric emission is strongly enhanced in plage and network features overlaying photospheric faculae and {network}. It follows that chromospheric indices such as the 10.7 cm radio flux \citep[$\fra$,][]{tapping13}, Ca II K $1\ \AA{}$ emission index \citep{bertello16}, Lyman $\alpha$ irradiance \citep{woods00}, and Mg II index \citep{heath86,snow14} are modulated with plage and chromospheric network emission, and their relation to sunspot indices offers another avenue to probe how faculae and {network} relate to sunspots. In this article, we present such an effort, examining the relationship between the aforementioned chromospheric indices and the group sunspot number, international sunspot number and total sunspot area. At the same time, we also look at how the total magnetic flux enclosed in faculae and {network} as apparent in full-disc solar magnetograms \citep{yeo14a}, denoted $\fmf$, relates to the various sunspot indices. The aim is to {complement} and extend the studies on the relationship between sunspot and facular area \citep[i.e.][]{foukal93,foukal96,foukal98,chapman97,shapiro14} by investigating how chromospheric emission and $\fmf$ relate to sunspots. The advantage is that while it is still a challenge to compare sunspot and facular area over multiple solar cycles, the $\fra$ goes back to 1947, and the other chromospheric indices and $\fmf$ to the 1970s (see Sect. \ref{data}), lending themselves to a multi-cycle comparison to sunspot indices. {It is worth pointing out that there are sources of variability in chromospheric and coronal emission other than their enhancement over faculae and {network}. For example, the enhancement of solar 10.7 cm emission in compact sources associated with sunspots and in coronal loops \citep{tapping87}. The $\fra$ and the various chromospheric indices are strongly, but not solely, modulated by faculae and network prevalence.}

How faculae and {network} relate to sunspots is complex in that the amount of faculae and {network} present at a particular time is not indicated by prevailing sunspots alone. Active regions and their decay products dissipate slower than the sunspots they bear, with the result being that the magnetic flux of a given active region persists, manifesting as faculae and network, even after the embedded sunspots have decayed \citep{vandrielgesztelyi15}. {This means, at a given time, there can be faculae and {network} present that are not associated with the sunspot-bearing active regions present {at} that moment, but with earlier active regions where the sunspots have already dissipated.} Also, magnetic flux emerges on the solar surface, not just in active regions, but also in ephemeral regions \citep{harvey93,harvey00} and in the form of the internetwork magnetic field \citep{livingston75,borrero17}. Ephemeral regions and the internetwork magnetic field contribute to the magnetic network, but as they do not contain sunspots, their prevalence is not captured by monitoring sunspots. {The internetwork magnetic field does not appear to vary over the solar cycle \citep{buehler13,lites14}, suggesting that its contribution to the magnetic network is invariant over cycle timescales. In contrast, the number of ephemeral regions varies along with the solar cycle \citep[i.e. higher at cycle maxima and lower at minima,][]{harvey93,harvey00}, indicating that their contribution to the magnetic network might similarly exhibit cyclic variability.}

\cite{preminger05,preminger06a,preminger06b,preminger07}, denoted here as PW, sought to reproduce solar irradiance, total photospheric magnetic flux, and various chromospheric and coronal indices, referred to here as the target indices, from total sunspot area by convolving it with the appropriate finite impulse response (FIR) filter.  In other words, they modelled the relationship between total sunspot area and the target indices as a linear transformation. For a given target index, the FIR filter is given by the deconvolution of total sunspot area from the target index. As such, the FIR filter encapsulates the response of the target index to sunspots. {The FIR filters indicate that the appearance of a sunspot would produce a response in the target indices over multiple rotation periods, and the time variation in this response is consistent with what is expected from the fact that active region magnetic flux persists, in the form of faculae and network, for some time after their sunspots have dissipated (c.f. Sect. \ref{modelpw}).} The application of the FIR filters to total sunspot area closely replicated the target indices, leading the authors to conclude that the relationship between total sunspot area and the various target indices is well represented by a linear transformation and {does not change} with time. A modification of the PW model was recently proposed by \cite{dudokdewitt18}, which we discuss in Sect. \ref{modelysk}.

The more recent investigations by \cite{svalgaard10}, \cite{tapping11}, \cite{livingston12}, and \cite{tapping17} reported that the relationship between the $\fra$ and the international sunspot number and total sunspot area appears to have been changing since solar cycle 23. {In each of these studies, the authors modelled the $\fra$ implicitly assuming the level at a particular time is a function of prevailing sunspots alone.} For example, \cite{tapping11} and \cite{tapping17}, hereinafter referred to collectively as TVM, described the $\fra$ as an exponential-polynomial function of the sunspot indices, described here in Sect. \ref{modeltvm}. {\cite{svalgaard10} also examined the scatter plot of the $\fra$ and the international sunspot number (see Fig. 2 in their paper). This compares the $\fra$ {at each time} to the sunspot number at {that} time, which again implicitly assumes that the $\fra$ is a function of prevailing sunspots alone. We had noted that due to the way active regions evolve, the amount of faculae and network present at a given time {is} indicated not just by the sunspots {present} at that moment, but also by sunspots in the recent past. This would, of course, extend to the response of the $\fra$ to faculae and network. The analyses of \cite{svalgaard10}, \cite{livingston12}, and TVM, by treating the F10.7 as a function of prevailing sunspots alone, does not take this into account. Since solar 10.7 cm emission is enhanced not just over sunspots, but also over faculae and network, it is inconclusive if the findings of these authors point to secular variability in the relationship between the $\fra$ and sunspots.}

{Let us} refer to chromospheric indices and $\fmf$, which are both modulated by faculae and network prevalence, collectively as facular indices. We examined the relationship between sunspot and facular indices through an empirical model that extends the linear transformation approach put forward by PW. In the following, we describe the sunspot and facular indices considered (Sect. \ref{data}) before presenting the model (Sect. \ref{model}). The realism of any model of the relationship between sunspot and facular indices is, of course, indicated by how well the facular indices can be reconstructed from the sunspot indices with the model. We demonstrate the proposed model to be competent in this regard (Sect. \ref{analysis1}). Making use of the model reconstruction of the facular indices from the sunspot indices and an empirical model of solar irradiance variability based on it, we examine how chromospheric emission and $\fmf$ scale with sunspots (Sect. \ref{analysis2}), and what the apparent relationship implies for solar and stellar brightness variations (Sect. \ref{analysis3}). Finally, we provide a summary of the study in Sect. \ref{summary}.

\section{Data}
\label{data}

\begin{figure}{h}
\centering
\resizebox{\hsize}{!}{\includegraphics{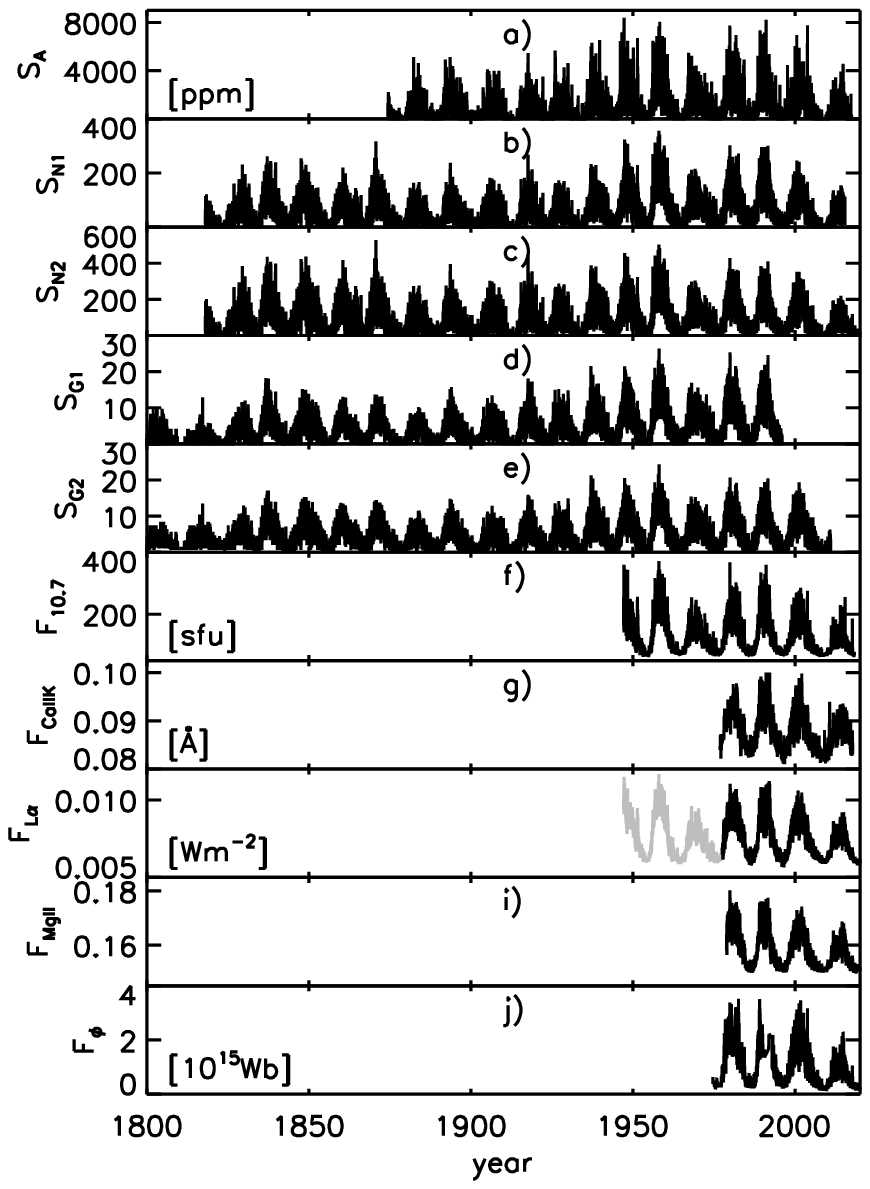}}
\caption{Sunspot and facular indices examined in this study. From top to bottom, total sunspot area ($\sa$), international sunspot number (original; $\snone$, revision; $\sntwo$), group sunspot number (original; $\sgone$, revision by \citealt{chatzistergos17}; $\sgtwo$), 10.7 cm radio flux ($\fra$), Ca II K $1\ \AA{}$ emission index ($\fca$), Lyman $\alpha$ irradiance ($\fla$), Mg II index ($\fmg$), and the total magnetic flux enclosed in faculae and network ($\fmf$). The $\sa$, $\fra$, $\fca$, $\fla$ and $\fmf$ are in units of ppm of the solar hemisphere, solar flux units, \r{A}ngstr\"{o}m, $\rm Wm^{-2}$ and Weber, respectively. The grey segment of the $\fla$ time series is excluded from the analysis. See Sect. \ref{data} for details.}
\label{fsrdata}
\end{figure}

In this study, we investigate how faculae and network relate to sunspots by examining the relationship between sunspot indices and what we term facular indices (defined in Sect. \ref{introduction}), denoted $S$ and $F$, respectively. Specifically, we compared {the daily total sunspot area, $\sa$, international sunspot number, $\sn$ and group sunspot number $\sg$ to the daily 10.7 cm radio flux, $\fra$, Ca II K $1\ \AA{}$ emission index, $\fca$, Lyman $\alpha$ irradiance, $\fla$, Mg II index, $\fmg$ and total magnetic flux enclosed in faculae and network, $\fmf$ (depicted in Fig. \ref{fsrdata}).} We made use of the Penticton $\fra$ record \citep{tapping13} and the composite time series of $\sa$ by \cite{balmaceda09}, $\fca$ by \cite{bertello16}, $\fla$ by \cite{machol19} and $\fmg$ provided by IUP (Institut f\"ur Umweltphysik, Universit\"at Bremen). The $\fmf$ time series is taken from \cite{yeo14a}, who isolated the faculae and network features in daily full-disc magnetograms dating back to 1974. While the \cite{machol19} $\fla$ composite goes back to 1947, the 1947 to 1977 segment (grey, Fig. \ref{fsrdata}h) is not provided by $\fla$ measurements but a model based on the 10.7 cm and 30 cm radio flux. For this reason, we exclude it from further consideration.

Both the $\sg$ and $\sn$ have been revised recently. Several competing revisions of the \cite{hoyt98} $\sg$ composite are available, namely by \cite{svalgaard16b}, \cite{usoskin16}, \cite{cliver16}, and \cite{chatzistergos17}. Of these, only the \cite{chatzistergos17} revision is suitable for the current study. We are interested in the daily $\sg$, but the \cite{svalgaard16b} revision is only available at monthly cadence due to the calibration method. The \cite{usoskin16} revision is moot here since it does not introduce any modifications to the \cite{hoyt98} time series after 1900, and therefore it compares similarly to the various facular indices as the original time series. As the \cite{cliver16} revision only goes up to 1976, we cannot compare it to the $\fca$, $\fla$ and $\fmg$ as none of these go further back than 1976. To find any effect of the changes introduced to the $\sg$ on the analysis, we looked at both the \cite{hoyt98} and \cite{chatzistergos17} time series, distinguished here as $\sgone$ and $\sgtwo$. For the same reason, we examined both the original and revised $\sn$ composites \citep{clette14,clette16b,clette16a}, denoted $\snone$ and $\sntwo$.

To investigate the implications of the apparent relationship between sunspot and facular indices on solar and stellar brightness variations, we also make use of the PMOD total solar irradiance (TSI) composite \citep[version $42\_65\_1709$,][]{frohlich00,frohlich06} and the \cite{balmaceda09} photometric sunspot index (PSI) composite. The PSI \citep{hudson82,frohlich94} indicates the proportional deficit in TSI due to sunspots.

We made use of the various data sets as available at the time of study, downloaded on 22 January 2020. The online sources are listed in the acknowledgements.

\section{Models}
\label{model}

We examined the relationship between the various sunspot and facular indices (Fig. \ref{fsrdata}) through an empirical model. The model, abbreviated to YSK, is an extension of the linear transformation approach proposed by PW \citep{preminger05,preminger06a,preminger06b,preminger07}. In Sect. \ref{analysis1}, we examine how well we can replicate the facular indices {from} the sunspot indices with the YSK model {and} with the PW and TVM \citep{tapping11,tapping17} approaches {(serving as control)}. Before that, we first describe the TVM (Sect. \ref{modeltvm}), PW (Sect. \ref{modelpw}), and YSK models (Sect. \ref{modelysk}). We denote the TVM model of facular index $F$ as a function of sunspot index $S$ as $\mfstvm$, and similarly that by PW and YSK as $\mfspw$ and $\mfsysk$. 

\subsection{TVM model}
\label{modeltvm}

{\cite{tapping11} and \cite{tapping17}} examined the relationship between the $\fra$ and the $\sa$ and $\sn$. They smoothed the various time series and fit an exponential-polynomial relationship of the form
\begin{equation}
\label{eqntvm}
\mfstvm=\left(2-\exp\left(-f_1{}S\right)\right)\left(f_2{}S^2+f_3{}S\right)+f_4,
\end{equation}
where $f_1$ to $f_3$ are fit parameters, with the condition that $f_1\geq0$, and $f_4$ is fixed at 67 sfu. In the comparison to the $\sn$, $f_2$ was also fixed at null. Here, we applied the TVM model (Equation \ref{eqntvm}) to the $S$ and $F$ data sets (Sect. \ref{data}) as they are (i.e. no smoothing) and without any constraints on $f_1$ to $f_4$, apart from the $f_1\geq0$ condition.

\subsection{PW model}
\label{modelpw}

\begin{figure}
\centering
\resizebox{\hsize}{!}{\includegraphics{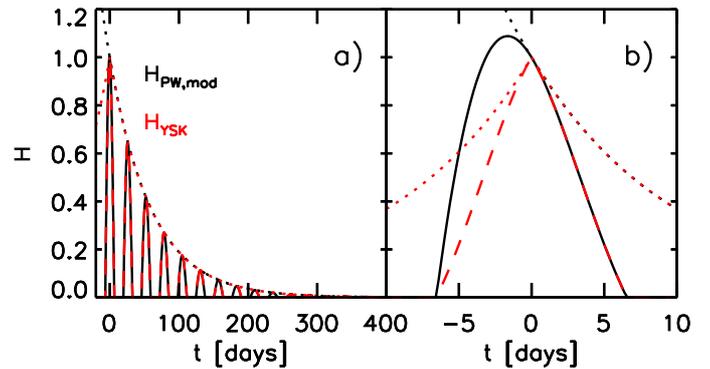}}
\caption{The model FIR filter in the PW model, $\hpwmod$ (black solid lines) and in the current model, $\hysk$ (red dashed lines) for time constants of a) 60 days and b) 10 days. The black and red dotted lines follow the exponential envelope of the respective model FIR filters. In the right panel, we limit the plot range to $-10\leq{}t\leq{}10$ days to highlight the difference in the two model FIR filters from the different envelope functions assumed. See Sect. \ref{model} for details.}
\label{fsrfir}
\end{figure}

Preminger \& Walton aimed to reproduce various chromospheric indices, including the four examined here, from the $\sa$. The response of a given chromospheric index to $\sa$ is modelled as a linear transformation. Specifically, as the convolution of $\sa$ with the finite impulse response (FIR) filter, $\hpwemp$ derived empirically by the deconvolution of $\sa$ from the chromospheric index. That is,
\begin{equation}
\label{eqnpw1}
\mfspwemp=\sa\otimes\hpwemp+g_3,
\end{equation}
where $g_3$ is a fit parameter. Preminger \& Walton found $\hpwemp$ to be consistent with what is expected from active region evolution. They described the form of $\hpwemp$ with a model FIR filter, $\hpwmod$ \citep{preminger07}. As a function of time, $t$:
\begin{equation}
\label{eqnpw2}
\hpwmod\left(t\right)=\max\left[\exp\left(-\frac{t}{g_2}\right)\cos\left(\frac{2\pi{}t}{t_\sun}\right),0\right],
\end{equation}
where $-0.25t_\sun{}\leq{}t\leq{}14.25t_\sun$ and $t_\sun$ is the synodic rotation period of the Sun, taken here to be 26.24 days. As illustrated in Fig. \ref{fsrfir}a, this describes a sequence of lobes of diminishing amplitude, as modulated by the exponential envelope of time constant $g_2$. The lobes are $0.5t_\sun$ wide and come in intervals of $t_\sun$, with the first lobe centred on $t=0$. The sequence of lobes is truncated at $t=14.25t_\sun$ on the observation that there is no discernible signal in $\hpwemp$ above this limit. Active regions emerge rapidly (days) and decay slowly (weeks to months). The response of $F$ to active regions as they emerge is represented by the rising edge of the first lobe. As active regions decay, the effect on $F$ not only diminishes with time but is also modulated by solar rotation, delineated by the sequence of lobes of diminishing amplitude. We note here that $\hpwmod$, defined such that it is fixed at unity at $t=0$, describes the form, but not the amplitude of $\hpwemp$.

While the empirical FIR filters comply with what is expected from active region evolution, they do not contain any features that can be clearly attributed to ephemeral regions. This {is the case} even though it is known that the number of ephemeral regions varies roughly in phase with the sunspot cycle \citep{harvey93,harvey00}, which alludes to a connection between ephemeral regions and sunspots. We surmise that though both sunspots and ephemeral regions are manifestations of the same magnetic cycle, ephemeral region emergence is not coupled to sunspot emergence. As such, the PW model is essentially a model of the effect of active regions on $F$.

Preminger \& Walton compared $\sa\otimes\hpwmod$, with $g_2$ fixed at certain arbitrary values, to the $\fra$ and Ap index \citep{foukal96}. Otherwise, there was no attempt to model the relationship between chromospheric and sunspot indices with $\hpwmod$. Generalising the PW model (Equation \ref{eqnpw1}) to other sunspot indices and taking into account that $\hpwmod$ describes the time-dependence of $\hpwemp$, we can rewrite Equation \ref{eqnpw1} as
\begin{equation}
\label{eqnpw3}
\mfspw=g_1{}S\otimes\max\left[\exp\left(-\frac{t}{g_2}\right)\cos\left(\frac{2\pi{}t}{t_\sun}\right),0\right]+g_3,
\end{equation}
where $g_1$ to $g_3$ are fit parameters. The $g_1$ term, missing from Equation \ref{eqnpw2}, scales the amplitude of the model FIR filter to the appropriate level. In our analysis, we applied this form of the PW model to the $S$ and $F$ data sets (Sect. \ref{data}).

\subsection{YSK model}
\label{modelysk}

The current model {is} an extension of the PW model (Equation \ref{eqnpw3}){. It} is given by
\begin{equation}
\label{eqnysk1}
\mfsysk=h_1{}S^{h_2}\otimes\hysk+h_4,
\end{equation}
where
\begin{equation}
\label{eqnysk2}
\hysk=\max\left[\exp\left(-\frac{|t|}{h_3}\right)\cos\left(\frac{2\pi{}t}{t_\sun}\right),0\right]
,\end{equation}
and $h_1$ to $h_4$ are fit parameters.

The model FIR filter here, $\hysk$ (Equation \ref{eqnysk2}) is identical to that in the PW model (Equation \ref{eqnpw2}), except the exponential envelope is given by $\exp{\left(-|t|/h_3\right)}$ instead of $\exp{\left(-t/h_3\right)}$. The effect on the model FIR filter is illustrated in Fig. \ref{fsrfir}b. While the envelope function adopted by PW skews the first lobe towards the negative time domain (black solid line), the proposed envelope function renders it symmetrical at about $t=0$ (red dashed line). We introduced this modification on the observation that the empirical FIR filters derived by PW \citep[see, for example, Fig. 1 in][]{preminger06a} do not indicate any clear skewness in the first lobe at about $t=0$.

In another departure from the PW model, the model FIR filter is applied to $S^{h_2}$ instead of $S$. By applying the FIR filter to $S$, the PW model implicitly assumes that, active region evolution aside, $F$ scales linearly with $S$, which is unlikely to be the case. It is known that chromospheric emission does not scale linearly with photospheric magnetic flux density \citep[e.g.][see also Sect. \ref{analysis2}]{schrijver89,harvey99,loukitcheva09,kahil17,barczynski18} and facular area is a quadratic function of sunspot area \citep{foukal93,foukal96,foukal98,chapman97,shapiro14}. This alludes to a non-linear relationship between plage emission and facular magnetic flux, and between facular magnetic flux and $S$. We introduced the $h_2$ parameter to take this into account.

As noted in the introduction, \cite{dudokdewitt18} presented a modified version of the PW model. In their model, the variables are at 27-day (instead of daily) cadence so as to exclude solar rotation effects from the FIR filter. In addition, the $g_3$ term (Equation \ref{eqnpw1}) is allowed to vary with time. The authors found that with the coarser time resolution, most of the long-term (annual to decadal) variation in $F$ is captured in $g_{3}(t)$ instead of the convolution of $\sa$ and the FIR filter. We did not adopt these modifications, as neither excluding solar rotation effects from the FIR filter nor splitting the variability in $F$ into two separate terms is necessary for the purposes of the current study.

\section{Analysis}
\label{analysis}

\subsection{Model validation}
\label{analysis1}

\begin{table*}
\caption{For each combination of $F$ and $S$, the fit parameters of the YSK model ($h_1$ to $h_4$, Equations \ref{eqnysk1} and \ref{eqnysk2}) are listed. See Sect. \ref{analysis1} for details.}
\label{yskptable}
\centering
\begin{tabular}{cccccc}
\hline\hline
 & & & & & \\
$F$ & $S$ & $h_1$ & $h_2$ & $h_3$ & $h_4$ \\
\hline
$\fra$ & $\sa$    & $1.369\times10^{-2}$ & 0.8603 & 24.35 & 64.38                \\
$\fca$ & $\sa$    & $5.030\times10^{-6}$ & 0.5854 & 57.41 & $8.228\times10^{-2}$ \\
$\fla$ & $\sa$    & $7.198\times10^{-7}$ & 0.6642 & 72.34 & $5.852\times10^{-3}$ \\
$\fmg$ & $\sa$    & $3.256\times10^{-6}$ & 0.7129 & 54.10 & 0.1497               \\
$\fmf$ & $\sa$    & $1.992\times10^{-4}$ & 0.8429 & 33.46 & 0.1824               \\
\hline
$\fra$ & $\snone$ & $7.334\times10^{-2}$ & 1.073  & 13.41 & 65.10                \\
$\fca$ & $\snone$ & $1.357\times10^{-5}$ & 0.7913 & 34.84 & $8.243\times10^{-2}$ \\
$\fla$ & $\snone$ & $3.230\times10^{-6}$ & 0.7960 & 49.25 & $5.861\times10^{-3}$ \\
$\fmg$ & $\snone$ & $1.614\times10^{-5}$ & 0.8648 & 33.25 & 0.1498               \\
$\fmf$ & $\snone$ & $1.735\times10^{-3}$ & 0.9572 & 20.15 & 0.1401               \\
\hline
$\fra$ & $\sntwo$ & $4.573\times10^{-2}$ & 1.092  & 13.42 & 65.44                \\
$\fca$ & $\sntwo$ & $1.078\times10^{-5}$ & 0.7846 & 34.52 & $8.254\times10^{-2}$ \\
$\fla$ & $\sntwo$ & $2.082\times10^{-6}$ & 0.8346 & 46.47 & $5.898\times10^{-3}$ \\
$\fmg$ & $\sntwo$ & $8.884\times10^{-6}$ & 0.9209 & 32.71 & 0.1501               \\
$\fmf$ & $\sntwo$ & $1.054\times10^{-3}$ & 0.9841 & 20.26 & 0.1401               \\
\hline
$\fra$ & $\sgone$ & 0.7818               & 1.126  & 19.53 & 64.01                \\
$\fca$ & $\sgone$ & $5.693\times10^{-5}$ & 0.9559 & 35.56 & $8.336\times10^{-2}$ \\
$\fla$ & $\sgone$ & $1.501\times10^{-5}$ & 0.9295 & 54.62 & $5.887\times10^{-3}$ \\
$\fmg$ & $\sgone$ & $1.156\times10^{-4}$ & 0.9539 & 31.37 & 0.1495               \\
$\fmf$ & $\sgone$ & $1.590\times10^{-2}$ & 0.9538 & 23.28 & 0.1898               \\
\hline
$\fra$ & $\sgtwo$ & 0.6196               & 1.255  & 19.68 & 63.91                \\
$\fca$ & $\sgtwo$ & $7.421\times10^{-5}$ & 0.9240 & 36.82 & $8.218\times10^{-2}$ \\
$\fla$ & $\sgtwo$ & $1.564\times10^{-5}$ & 1.002  & 48.40 & $5.850\times10^{-3}$ \\
$\fmg$ & $\sgtwo$ & $8.279\times10^{-5}$ & 1.105  & 34.04 & 0.1499               \\
$\fmf$ & $\sgtwo$ & $1.361\times10^{-2}$ & 1.072  & 23.83 & 0.1279               \\
\hline
\end{tabular}
\end{table*}

\begin{table*}
\caption{For each combination of $F$ and $S$, the agreement between $F$ and the reconstruction from the TVM, PW and YSK models, as indicated by $\fittwo$, $\fitone$ and $\fitthree$, defined in Sect. \ref{analysis1}.}
\label{fittable}
\centering
\begin{tabular}{ccccccccccc}
\hline\hline
 & & & & & & & & & & \\
 & & \multicolumn{3}{c}{$\fittwo$ $[10^{-2}]$} & \multicolumn{3}{c}{$\fitone$ $[10^{-3}]$} & \multicolumn{3}{c}{$\fitthree$ $[10^{-2}]$}\\ 
$F$ & $S$ & TVM & PW & YSK & TVM & PW & YSK & TVM & PW & YSK \\
\hline
$\fra$ & $\sa$    & 14.30 & 6.453 & 6.036 & 20.01 & 8.493 & 7.904 & 6.599 & 5.797 & 5.949 \\
$\fca$ & $\sa$    & 35.54 & 26.87 & 17.62 & 60.57 & 42.26 & 25.37 & 13.91 & 12.67 & 7.659 \\
$\fla$ & $\sa$    & 31.09 & 20.15 & 10.78 & 50.65 & 29.87 & 14.66 & 12.23 & 10.18 & 6.209 \\
$\fmg$ & $\sa$    & 26.55 & 17.17 & 9.283 & 41.50 & 24.83 & 12.47 & 11.93 & 9.848 & 6.799 \\
$\fmf$ & $\sa$    & 20.84 & 10.44 & 9.728 & 30.88 & 14.19 & 13.25 & 10.52 & 6.321 & 6.390 \\       
\hline
$\fra$ & $\snone$ & 9.779 & 6.917 & 6.767 & 13.20 & 9.136 & 8.922 & 5.773 & 4.901 & 4.972 \\
$\fca$ & $\snone$ & 20.06 & 14.27 & 13.45 & 29.52 & 19.99 & 18.70 & 7.242 & 7.084 & 5.224 \\
$\fla$ & $\snone$ & 17.92 & 12.18 & 9.659 & 25.87 & 16.83 & 13.02 & 8.401 & 8.831 & 6.634 \\
$\fmg$ & $\snone$ & 13.82 & 7.689 & 7.375 & 19.28 & 10.21 & 9.759 & 7.335 & 7.376 & 6.531 \\
$\fmf$ & $\snone$ & 13.56 & 9.628 & 9.587 & 18.87 & 13.01 & 12.78 & 7.428 & 6.050 & 6.138 \\       
\hline
$\fra$ & $\sntwo$ & 9.235 & 6.383 & 6.186 & 12.40 & 8.395 & 8.109 & 9.675 & 5.466 & 4.929 \\
$\fca$ & $\sntwo$ & 21.71 & 15.57 & 14.79 & 32.42 & 22.05 & 20.79 & 8.605 & 8.134 & 6.679 \\
$\fla$ & $\sntwo$ & 15.57 & 9.704 & 7.634 & 22.04 & 13.13 & 10.12 & 6.868 & 6.716 & 4.856 \\
$\fmg$ & $\sntwo$ & 12.12 & 5.990 & 5.879 & 16.67 & 7.850 & 7.688 & 5.683 & 4.988 & 4.572 \\
$\fmf$ & $\sntwo$ & 11.29 & 7.405 & 7.370 & 15.42 & 9.826 & 9.638 & 5.450 & 3.851 & 3.963 \\       
\hline
$\fra$ & $\sgone$ & 17.86 & 13.12 & 12.93 & 25.78 & 18.21 & 17.90 & 6.817 & 4.603 & 3.418 \\
$\fca$ & $\sgone$ & 24.17 & 17.88 & 17.86 & 36.90 & 25.83 & 25.77 & 6.526 & 4.872 & 4.833 \\
$\fla$ & $\sgone$ & 19.76 & 11.75 & 11.70 & 29.00 & 16.11 & 16.04 & 7.934 & 4.455 & 4.484 \\
$\fmg$ & $\sgone$ & 14.85 & 8.882 & 8.848 & 20.90 & 11.91 & 11.85 & 5.555 & 3.544 & 3.399 \\
$\fmf$ & $\sgone$ & 18.73 & 13.49 & 13.36 & 27.24 & 18.81 & 18.83 & 4.126 & 2.416 & 1.996 \\
\hline
$\fra$ & $\sgtwo$ & 17.19 & 12.42 & 11.87 & 24.67 & 17.13 & 16.29 & 6.927 & 5.373 & 4.030 \\
$\fca$ & $\sgtwo$ & 20.45 & 14.08 & 14.03 & 30.20 & 19.69 & 19.61 & 7.921 & 6.225 & 4.382 \\
$\fla$ & $\sgtwo$ & 16.56 & 8.561 & 8.557 & 23.63 & 11.44 & 11.43 & 6.823 & 4.142 & 4.176 \\
$\fmg$ & $\sgtwo$ & 13.39 & 6.890 & 6.801 & 18.61 & 9.090 & 8.958 & 5.518 & 3.244 & 2.929 \\
$\fmf$ & $\sgtwo$ & 14.65 & 9.811 & 9.785 & 20.57 & 13.26 & 13.25 & 4.433 & 2.005 & 2.027 \\
\hline
\end{tabular}
\end{table*}

We modelled the relationship between each facular index, $F$ and each sunspot index, $S$. Taking each combination of $F$ and $S$, we fitted the YSK model (Equations \ref{eqnysk1} and \ref{eqnysk2}) and the TVM (Equation \ref{eqntvm}) and PW models (Equation \ref{eqnpw3}){, (serving as control)}. The fit parameters of the YSK model are listed in Table \ref{yskptable}.

\begin{figure}[H]
\centering
\resizebox{8cm}{!}{\includegraphics{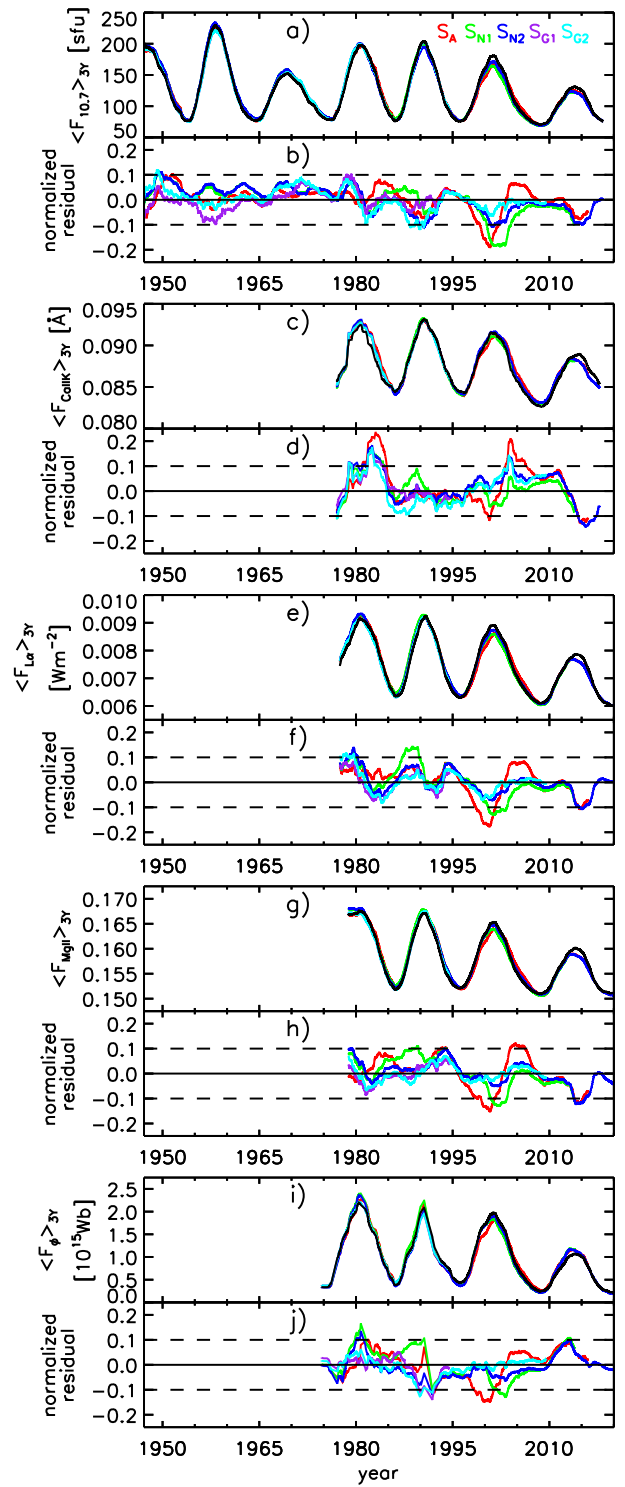}}
\caption{a) Three-year running mean of $\fra$ (black) and the YSK model reconstruction of this facular index from the $\sa$ (red), $\snone$ (green), $\sntwo$ (blue), $\sgone$ (purple), and $\sgtwo$ (cyan). b) Residual between the observed and reconstructed time series, normalised to the change in the former between the 2000 solar cycle maximum and 2008 minimum (Equation \ref{eqnnr}). The dashed lines mark the $10\%$ bound. c-j) The corresponding plots for the $\fca$, $\fla$, $\fmg$ and $\fmf$.}
\label{fsrstability}
\end{figure}

To validate the YSK model, we examined how well it reproduces the facular indices from the sunspot indices as compared to the two control models. For each model and each combination of $F$ and $S$, we derive the following. {We calculate the deviation from unity of the Pearson's correlation coefficient between $F$ and the model reconstruction of $F$ from $S$, $\mfs$, denoted $\fittwo$. This quantity indicates the variability in $F$ that is not replicated in $\mfs$. Taking the $F$-versus-$\mfs$ scatter plot, we derive the ratio of the variance normal to and in the direction of the $F=\mfs$ line, denoted as $\fitone$. The more $\mfs$ replicates the variability and the scale of $F$, the lower the value of $\fitone$. To reveal how closely the long-term (annual to decadal) trend in $F$ is reproduced in $\mfs$, we took the three-year running mean of $F$ and $\mfs$, denoted as $\smoothf$ and $\smoothmfs$, and calculate the normalised residual, given by
\begin{equation}
\label{eqnnr}
\frac{\smoothf-\smoothmfs}{\smoothfmax-\smoothfmin}.
\end{equation}
This is the difference between $\smoothf$ and $\smoothmfs$, normalised to the change in the former between the 2000 solar cycle maximum and 2008 minimum. Following convention, the epoch of solar cycle extrema is taken from the 13-month moving mean of the monthly $\sntwo$. The normalisation expresses the residual as a proportion of solar cycle variability. The discrepancy between the measured and modelled long-term variability is also encapsulated in the root-mean-square of the normalised residual, abbreviated as $\fitthree$.} We tabulate $\fittwo$, $\fitone$ and $\fitthree$ in Table \ref{fittable}, and depict the normalised residue of the YSK model in Fig. \ref{fsrstability}.

For the YSK model, $\fittwo$ ranges from about 0.06 to 0.18 and $\fitthree$ from 0.02 to 0.08 (Table \ref{fittable}), indicating that it reproduces about $82\%$ to $94\%$ of the variability in the various facular indices and their long-term trend to about $2\%$ to $8\%$ of solar cycle variability. In terms of $\fittwo$, $\fitone$ and $\fitthree$, the YSK and PW models replicate the facular indices better than the TVM model. The only exceptions are the $\fmg$ \&\ $\snone$ and $\fla$ \&\ $\snone$ analyses, where only the YSK model registered a lower $\fitthree$ than the TVM model. The strength of the YSK and PW models over the TVM model highlights {how important it is, in such studies,} to account for the fact that active region magnetic fluxes decay slower than sunspots (as similarly argued by \citealt{foukal98} {and} \citealt{preminger07}), and the suitability of the linear transformation approach proposed by PW for this purpose. The PW model registered a lower $\fitone$ than the YSK model in the $\fca$ {\&} $\sntwo$ and $\fmf$ {\&} $\sgone$ analyses, and a lower $\fitthree$ for eight of the 25 combinations of $F$ and $S$. Otherwise, the YSK model achieved a lower value of $\fittwo$, $\fitone$ and $\fitthree$ than the PW model. Overall, the YSK model describes the relationship between the sunspot and facular indices better than both the TVM and PW models.

{Secular variation in the relationship between $F$ and $S$, if present, will imprint} itself on the normalised residual. This will, however, be obscured by fluctuations in the normalised residual from data uncertainty and model limitations. To count any trend in the normalised residual as corresponding to secular variability in the relationship between $F$ and $S$ with confidence, it has to be apparent in multiple combinations of $F$ and $S$, and significant compared to $\fitthree$. For the YSK model, in absolute terms, the normalised residual (Fig. \ref{fsrstability}) is below $10\%$ (dashed lines) almost everywhere, meaning it is comparable to $\fitthree$ ($2\%$ to $8\%$, Table \ref{fittable}). Looking at the $\fra$ {\&} $\sa$ (red, Fig. \ref{fsrstability}b), $\fra$ {\&} $\snone$ (green, Fig. \ref{fsrstability}b){,} and $\fla$ {\&} $\sa$ analyses (red, Fig. \ref{fsrstability}f), the normalised residual rose by up to $20\%$ between 2000 and 2005. However, this is not corroborated by how these two facular indices compare to the reconstruction from the other sunspot indices (Figs. \ref{fsrstability}b and \ref{fsrstability}f), or by the $\fca$ (Fig. \ref{fsrstability}d), $\fmg$ (Fig. \ref{fsrstability}h), and $\fmf$ analyses (Fig. \ref{fsrstability}j). In the case of the $\fca$ {\&} $\sa$ analysis (red, Fig. \ref{fsrstability}d), the normalised residual rose to about $20\%$ between 1980 and 1985, and again around 2005, but as before, this is not corroborated by any of the other combinations of $F$ and $S$. Within the limits of the current analysis, there is no clear evidence of any secular variation in the relationship between $F$ and $S$ over the past four decades. We remind the reader that \cite{svalgaard10}, \cite{livingston12}, and TVM found the relationship between the $\fra$ and the $\sa$ and $\sn$ to have been changing since solar cycle 23. The analysis here, making use of a model demonstrated to be more physical than the TVM model and extended to include more facular and sunspot indices, indicates otherwise. We conclude that the contradictory results from these earlier studies might be an artefact of data uncertainty and model limitations.

The $\sa$ is a continuous quantity, at least to the resolution limit of the underlying sunspot area measurements, but the $\sn$ and $\sg$, indicating the number of sunspots and sunspot groups, are discrete quantities. In the PW and YSK models, $F$ at a given time is, in effect, given by the weighted sum of $S$ over an extended period. The result is that the discrete nature of the $\sn$ and $\sg$ is not seen in the model reconstruction of $F$ from them. In other words, this time-averaging suppresses the effect of the quantization noise in $\sn$ and $\sg$ on the model output. It is not straightforward to isolate and quantify the uncertainty introduced into the YSK model by $\sn$ and $\sg$ being discrete, but the impact on the current discussion is likely to be minimal. As is evident from Fig. \ref{fsrstability}, at least in terms of the three-year running mean, the modelling results from $\sn$ (green and blue) and $\sg$ (purple and cyan) are consistent with that from $\sa$ (red).

\subsection{How faculae and {network} relate to sunspots}
\label{analysis2}

\begin{figure*}
\centering
\includegraphics{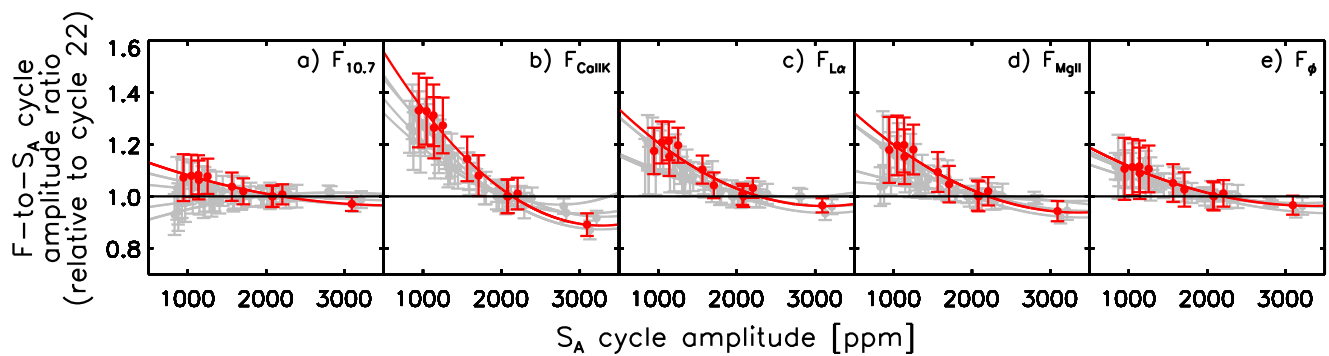}
\caption{Red: for each facular index, the $F$-to-$\sa$ solar cycle amplitude ratio, relative to cycle 22, as a function of $\sa$ cycle amplitude. The dots represent the individual cycle values, determined using the model reconstruction of $F$ from $\sa$. The error bars and curves denote the corresponding $1\sigma$ uncertainty and quadratic polynomial fit. Grey: the $F$-to-$S$ cycle amplitude ratio-versus-$S$ cycle amplitude profiles for the $\snone$, $\sntwo$, $\sgone$ and $\sgtwo$. The various sunspot indices are rescaled to the $\sa$ scale to allow a direct comparison. Only the values from cycles 12 to 22, where the five sunspot index time series overlap, are depicted. The horizontal line marks the cycle 22 level (i.e. unity). See Sect. \ref{analysis2} for details.}
\label{fsrcyclestrength}
\end{figure*}

As noted in the introduction, the {amount} of faculae and network present on the solar disc at a given time is not indicated by prevailing sunspots alone due to how active regions evolve and the contribution by ephemeral regions and the internetwork magnetic field. In the YSK model, $\mfsysk$, active region evolution is taken into account by the convolution of the sunspot indices, $S$ with the model FIR filter, $\hysk$ (Equations \ref{eqnysk1} and \ref{eqnysk2}). So whether in measurements or in the YSK model, a particular level of $S$ does not map to a unique value of $F$. Nonetheless, we can still gain insight into how faculae and network relate to sunspots by looking at the overall trend in $F$ with $S$. To this end, we compared the amplitude of the solar cycle in $\mfsysk$ and $S$. We opted to compare $\mfsysk$ instead of the measured $F$ to $S$ because of the following considerations. {Since the $S$ time series go further back in time than the $F$ time series (Fig. \ref{fsrdata}), and $\mfsysk$ evidently extend as far as $S$, we can compare $\mfsysk$ and $S$ over longer periods than when comparing $F$ and $S$. More critically, how the cycle amplitude in $F$ and $S$ compare can be affected by the uncertainty in the decadal trend in the various time series. This uncertainty is irrelevant when comparing $\mfsysk$ and $S$.} We recognise that the comparison of $\mfsysk$ and $S$ is only valid as far as the YSK model is physical, but the robustness of the model, demonstrated in Sect. \ref{analysis1}, renders confidence in this approach. The $\sa$ is a more direct measure of sunspot prevalence than the $\sn$ and $\sg$, which give sunspots and sunspot groups of different areas the same weighting. With this in mind, the focus here is on how $\mfsaysk$ and $\sa$ compare.

We use $\Delta{}F$ as an abbreviation {of} the deviation in $F$ from the 2008 solar cycle minimum level. Here, we define the cycle amplitude of $F$ and $S$, which we denote as $\af$ and $\as$, as the value of the three-year running mean of $\Delta{}F$ and $S$ at cycle maxima. In Fig. \ref{fsrcyclestrength}, we chart $\ratiosa$ against $\asa$ (red), revealing the trend in the $F$-to-$\sa$ cycle amplitude ratio with $\sa$ cycle amplitude as indicated by the YSK model. To compare the results from the various facular indices, we normalised the $F$-to-$\sa$ cycle amplitude ratio from each index to the cycle 22 value. The uncertainty in the $F$-to-$\sa$ cycle amplitude ratio, marked in the figure, is propagated from the uncertainty in the long-term trend in $\mfsysk$, $\fitthree$ (Table \ref{fittable}). For the various facular indices, the $F$-to-$\sa$ cycle amplitude ratio decreases with increasing $\sa$ cycle amplitude (Fig. \ref{fsrcyclestrength}). The decline is steepest for the $\fca$ (Fig. \ref{fsrcyclestrength}b), followed by the $\fla$ and $\fmg$ (where the trend with $\sa$ cycle amplitude is closely similar, Figs. \ref{fsrcyclestrength}c and \ref{fsrcyclestrength}d), then the $\fmf$ (Fig. \ref{fsrcyclestrength}e), and finally the $\fra$ (Fig. \ref{fsrcyclestrength}a). For the $\fra$, the trend is weak in relation to the uncertainty.

As a check, we repeated the above analysis with $\mfsnoneysk$ and $\snone$, that is, we examined $\ratiosnone$ as a function of $\asnone$. To allow a direct comparison to the $\mfsaysk$ and $\sa$ analysis, $A\left(\snone\right)$ is calculated after rescaling this sunspot index to the scale of the $\sa$ using the quadratic polynomial fit to the $\sa$-versus-$\snone$ scatter plot. This is repeated for $\mfsntwoysk$ and $\sntwo$, $\mfsgoneysk$ and $\sgone$, and $\mfsgtwoysk$ and $\sgtwo$. The results are drawn in grey in Fig. \ref{fsrcyclestrength}. The $\ratio$-versus-$A\left(S\right)$ profiles from the various sunspot indices lie largely within $1\sigma$ of one another, indicating that they are mutually consistent, affirming what we noted with the $\mfsaysk$ and $\sa$ analysis (red). Notably, for the $\fra$ (Fig. \ref{fsrcyclestrength}a), while the various profiles are within error of one another, they indicate conflicting trends with $\sa$ cycle amplitude. In other words, for this particular facular index, the underlying trend is too weak to be established by the current analysis. 

The observation here that the $\fra$ departs from the other facular indices in terms how it compares to the $\sa$ is at least partly due to the following. {We noted in the introduction that while the various chromospheric indices are strongly modulated by faculae and network prevalence due to the enhancement of chromopheric emission over these photospheric magnetic features, there are other sources of variability. Solar 10.7 cm emission is enhanced in compact sources that are associated with sunspots \citep{tapping87}. This is not the case for the Ca II H\&K, Lyman $\alpha$, and Mg II h\&k lines, and while sunspots can still be darker or brighter in these lines, depending on height in the solar atmosphere, the effect is much weaker. Clearly, this would have contributed to the divergence between the $\fra$ and the other chromospheric indices and $\fmf$ noted here.}

Excluding the $\fra$, the $F$-to-$\sa$ cycle amplitude ratio decreases more steeply with increasing $\sa$ cycle amplitude for the chromospheric indices (Figs. \ref{fsrcyclestrength}b to \ref{fsrcyclestrength}d) than for the $\fmf$ (Fig. \ref{fsrcyclestrength}e). This means the chromospheric index-to-$\fmf$ cycle amplitude ratio also diminishes with rising $\sa$ cycle amplitude. We attribute this to how chromospheric emission relates to the photospheric magnetic field. Various studies have noted that at chromospheric passbands, the relationship between the intensity excess of chromospheric features and the underlying photospheric magnetic flux density can be described by a power law with an exponent that is below unity \citep[e.g.][]{schrijver89,harvey99,loukitcheva09,kahil17,barczynski18,chatzistergos19b}. In fact, part of the purpose of the $h_2$ term in the YSK model (Equation \ref{eqnysk1}) is to capture this relationship (c.f. Sect. \ref{modelysk}). The power-law exponent lying below unity means the ratio of the intensity excess and photospheric magnetic flux density of chromospheric features declines with rising photospheric magnetic flux density. The $\fca$-to-$\fmf$ ratio decreasing with increasing $\sa$ cycle amplitude is the extension of this behaviour to the disc-integrated Ca II K emission and magnetic flux. The same argument applies to the $\fla$ and $\fmg$. Of course, the more the power-law exponent deviates from unity, the more pronounced this effect is.

The power-law exponent has been reported for the Ca II K \citep{schrijver89,harvey99,loukitcheva09,chatzistergos19b} and Mg II k lines \citep{barczynski18}. However, since the passband of the Mg II k filtergrams used in the \cite{barczynski18} study differs from the spectral sampling of the Mg II h\&k doublet in the derivation of the $\fmg$, the reported exponents are of limited relevance to the $\fmg$. The situation is similar for the $\fca$. To the best of our knowledge, the power-law exponent at the Lyman $\alpha$ line has not been reported in the literature. The study by \cite{barczynski18}, which examined six passbands formed at various heights in the solar atmosphere{,} noted the following. Going from the upper photosphere to the transition region, the power-law exponent decreases with height, up to the temperature minimum, {before it starts to increase with height instead}. The Mg II h\&k doublet is formed higher in the chromosphere than the Ca II K line \citep{leenaarts13a,leenaarts13b}, and the Lyman $\alpha$ line is formed even higher, {at the boundary to the} transition region \citep{vernazza81}. So while the power-law exponent at each line {as it would apply to the} corresponding chromospheric index is not known, the \cite{barczynski18} study and the formation height of the various lines do suggest that it should be lower for the Ca II K line than for the Lyman $\alpha$ and Mg II h\&k lines. This is consistent with the $F$-to-$\sa$ cycle amplitude ratio dropping more steeply with rising $\sa$ cycle amplitude for the $\fca$ (Fig. \ref{fsrcyclestrength}b) than for the $\fla$ (Fig. \ref{fsrcyclestrength}c) and $\fmg$ (Fig. \ref{fsrcyclestrength}d).

As stated in the introduction, various studies have found the facular-to-sunspot-area ratio to decrease with increasing sunspot area \citep{foukal93,foukal96,foukal98,chapman97,shapiro14}. In this section, we see that like facular area, the variation in chromospheric emission and the total magnetic flux enclosed in faculae and {network} scales with sunspots in such a way that the ratio {to} sunspot area also decreases with increasing sunspot area (Figs. \ref{fsrcyclestrength}b to \ref{fsrcyclestrength}e).

\subsection{Implication for solar and stellar brightness variations}
\label{analysis3}

\begin{figure}
\centering
\includegraphics{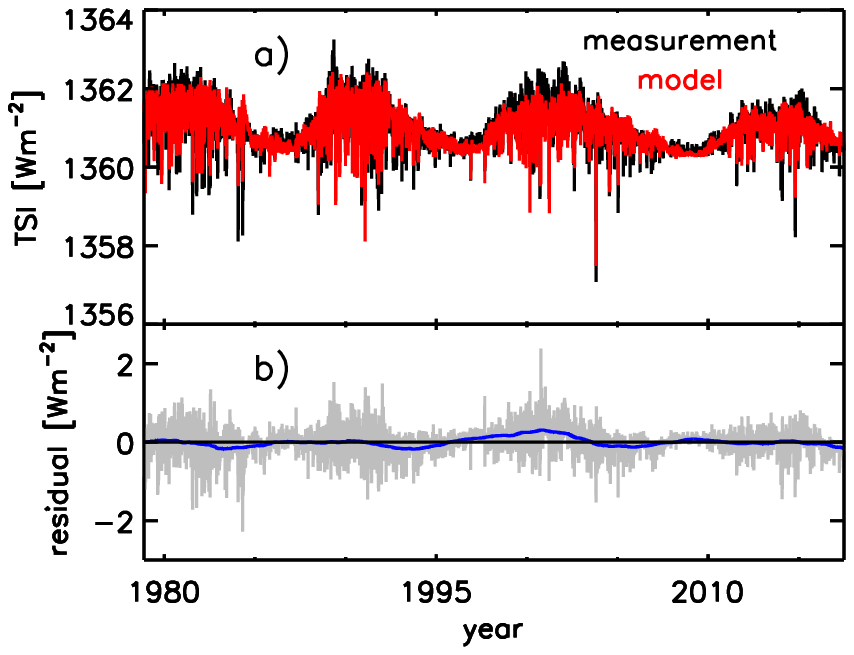}
\caption{a) PMOD TSI composite (black) and the model based on the multiple linear regression of the reconstruction of the $\fca$ from $\sa$, $\mfcasaysk$ and the PSI to this TSI time series (red, Equation \ref{eqntsi}). b) The difference between measurement and model (grey), and the corresponding three-year running mean (blue). The models based on $\mflasaysk$ and $\mfmgsaysk$ compare similarly to the PMOD TSI composite (Table \ref{tsitable}), not shown to avoid repetition.}
\label{fsrtsimodel}
\end{figure}

\begin{table*}
\caption{The fit parameters of the empirical TSI models ($k_1$ to $k_3$, Equation \ref{eqntsi}) derived taking the reconstruction of the $\fca$, $\fla$ and $\fmg$ from $\sa$, denoted $\mfcasaysk$, $\mflasaysk$ and $\mfmgsaysk$, as the FB proxy (i.e. the proxy of the TSI excess from faculae and network). The various models are optimised to the PMOD TSI composite in such a way that $k_2$ is identical. The agreement between model and measurement in terms of $\fittwo$, $\fitone$ and $\fitthree$, defined in Sect. \ref{analysis1}, is also tabulated.}
\label{tsitable}
\centering
\begin{tabular}{ccccccc}
\hline\hline
 & & & & & & \\
FB proxy & $k_1$ & $k_2$ & $k_3$ & $\fittwo$ & $\fitone$ $[10^{-2}]$ & $\fitthree$ \\
\hline
$\mfcasaysk$ & 209.0  & 10.36 & 1343.09 & 0.3072 & 4.954 & 0.1102 \\
$\mflasaysk$ & 643.8  & 10.36 & 1356.58 & 0.3014 & 4.785 & 0.1152 \\
$\mfmgsaysk$ & 119.1  & 10.36 & 1342.57 & 0.3088 & 5.173 & 0.1106 \\
\hline
\end{tabular}
\end{table*}
                        
\begin{figure}
\centering
\includegraphics{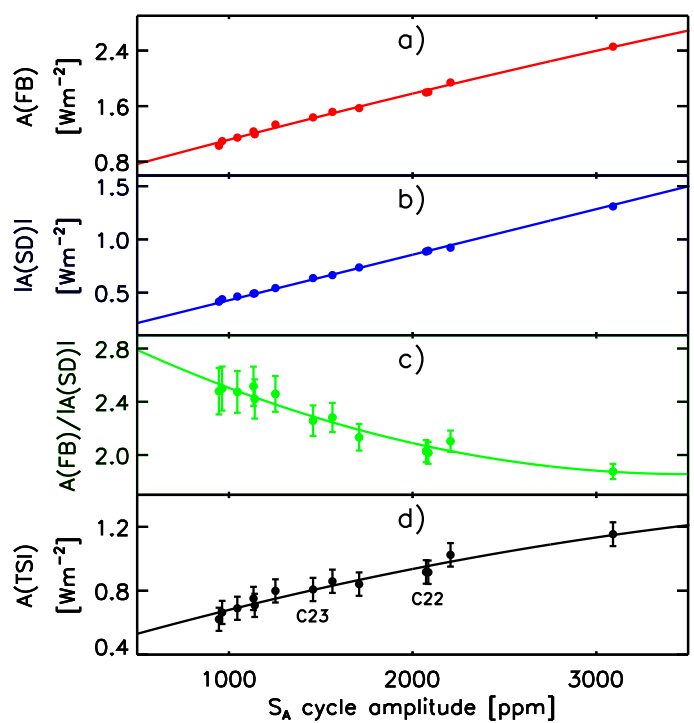}
\caption{From top to bottom, as a function of $\sa$ cycle amplitude, the cycle amplitude of a) the effect of faculae and network (red), and b) of sunspots on TSI (blue), c) the ratio of the two (green), and d) TSI (black). That is, $\afb$, $\left|\asd\right|$, $\afb/\left|\asd\right|$, and $\atsi$-versus-$\asa$. The plot points and error bars represent the mean of the values from the three TSI models and the associated $1\sigma$ uncertainty (see Sect. \ref{analysis3}). The latter is omitted in the case of $\afb$ and $\asd$, where it is so minute as to be obscured by the plot points. The values from the individual TSI models, lying within $1\sigma$ of the mean, are not drawn to avoid cluttering. The $\sa$ and PSI time series on which the TSI models are based, and therefore the plot points, cover solar cycles 12 to 24. The $\atsi$ plot points corresponding to solar cycles 22 and 23 are labelled. The lines correspond to the linear or quadratic polynomial fit.}
\label{fsrtsicyclestrength}
\end{figure}

\begin{figure}
\centering
\includegraphics{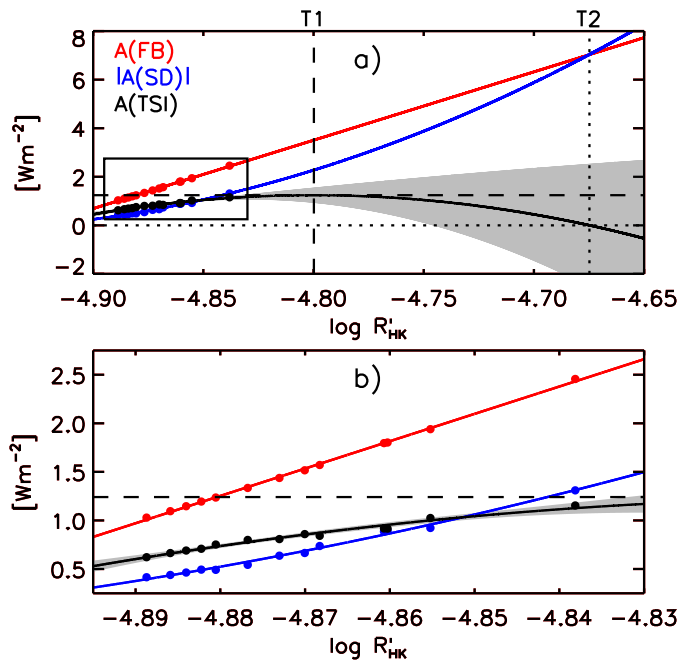}
\caption{Cycle amplitude of FB (red plot points), SD (blue) and TSI (black), taken from Fig. \ref{fsrtsicyclestrength}, as a function of $\logrhk$. The corresponding $1\sigma$ uncertainty, not drawn, is generally much smaller than the plot symbols. The red line follows the linear fit to $\afb$ and the blue line the quadratic fit to $\left|\asd\right|$, while the black line indicates the corresponding $\atsi$ level. {The shaded region encloses the $95\%$ confidence interval of the $\atsi$ curve.} The dashed lines mark the turning point of the $\atsi$ curve and the dotted lines where it goes below zero, denoted T1 and T2, respectively. See Sect. \ref{analysis3} for the physical interpretation. The boxed area is blown up in the bottom panel.}
\label{fsrstellar}
\end{figure}

\begin{figure}
\centering
\includegraphics{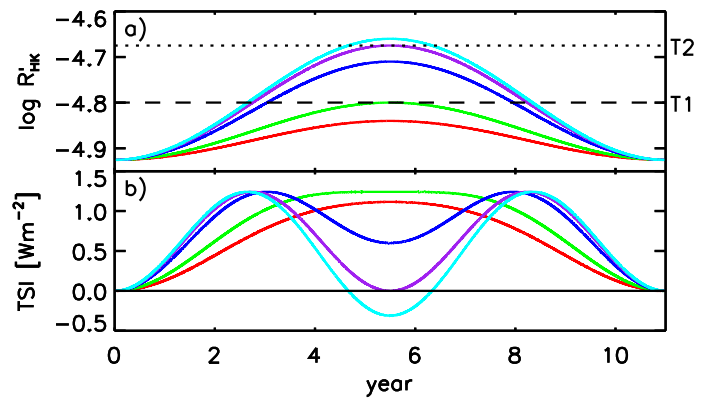}
\caption{a) The $\logrhk$ over a simulated solar cycle. The dashed and dotted lines, as in Fig. \ref{fsrstellar}, mark $\logrhk$ at T1 and T2. The scenario where the activity peak is below T1 (red), at T1 (green), between T1 and T2 (blue), at T2 (purple) and above T2 (cyan) are depicted. b) The TSI corresponding to each scenario, similarly colour-coded, as given by the empirical relationship between $\atsi$ and $\logrhk$ (black curve, Fig. \ref{fsrstellar}). The difference to the 2008 minimum TSI level (solid black line) is drawn.}
\label{fsrhypothetical}
\end{figure}

To examine the implications of the apparent relationship between facular and sunspot indices on solar and stellar brightness variations, we make use of the empirical model of this relationship, $\mfsysk$ to reconstruct the variation in TSI. For this purpose, we adopt what is termed the proxy approach \citep[e.g.][]{lean88,chapman96,chapman13,lean97,coddington16,yeo17a}. We refer to the TSI excess from faculae and {network} as facular brightening, and the deficit from sunspots as sunspot darkening, abbreviated as FB and SD, respectively. In the proxy approach, TSI variability is given by the multiple linear regression of a chromospheric index and a sunspot index, acting as proxies of FB and SD, to measured TSI. Here, we employ the reconstruction of $F$ from $\sa$, $\mfsaysk$ as the FB proxy and the $\ps$ as the SD proxy. {The $\ps$ indicates the proportional deficit in TSI due to sunspots \citep{hudson82,frohlich94}, and is calculated from the same sunspot area and position measurements as the $\sa$.} The TSI model is given by
\begin{equation}
\label{eqntsi}
\tsi=k_1{}\mfsaysk+k_2{}\ps+k_3,
\end{equation}
where $k_1$ to $k_3$ are fit parameters. {What we did here is incorporate the YSK model into an existing TSI modelling technique. The objective being to examine what the observations we made earlier, based on the YSK model{,} about the relationship between $F$ and $\sa$ (Sect. \ref{analysis2}) might imply for the effect of faculae, network, and sunspots on solar and stellar brightness variations. Using $\mfsaysk$ as the FB proxy also allowed us to reconstruct TSI over a longer period than if we used the significantly shorter measured $F$ time series.}

We derived three TSI models taking the reconstruction of the $\fca$, $\fla$ and $\fmg$ from $\sa$ as the FB proxy in turn. To minimise any bias from data uncertainty, the three models were optimised to the PMOD TSI composite in such a way that the $k_2$ term, the coefficient of the PSI in Equation \ref{eqntsi}, is common. In Fig. \ref{fsrtsimodel}a, we depict the model based on the $\fca$ reconstruction (red) along the PMOD composite (black). The fit parameters and the agreement between modelled and measured TSI, in terms of $\fittwo$, $\fitone$ and $\fitthree$ (defined in Sect. \ref{analysis1}), are summarised in Table \ref{tsitable}. Looking at $\fittwo$ and $\fitthree$, the TSI models replicate about $69\%$ of the variability in the PMOD composite and the long-term trend to about $11\%$ to $12\%$ of solar cycle variability (see also, Fig. \ref{fsrtsimodel}b). The three models replicate the PMOD composite reasonably well, and almost equally so. {For comparison, the TSI models by \cite{dasiespuig16} and \cite{pelt17}, {also} based on sunspot area and position measurements, reproduced about $76\%$ and $69\%$ of the variability in the PMOD composite, respectively.}

{We did not model TSI with the $\fra$ for the following reasons. The proxy approach requires a quantity that represents the effect of faculae and network on TSI. The enhancement of solar 10.7 cm emission over sunspots makes the $\fra$ less suited for this purpose than the $\fca$, $\fla$ and $\fmg$, where the intensity deficit or excess over sunspots provides a much smaller contribution to total emission (c.f. Sect. \ref{analysis2}). While the main contribution to TSI is from the photosphere, the Ca II H\&K and Mg II h\&k doublets are formed in the chromosphere, the Lyman $\alpha$ line at the boundary to the transition region, and solar 10.7 cm emission in the corona. The physical processes underlying how photospheric magnetism modulates the temperature structure and brightness of the solar atmosphere changes as we go up in height, such that the various indices would vary differently with faculae and {network as compared to} TSI. For example, in the lower photosphere, magnetic flux tubes are heated through their side walls by radiative heating, and in the upper photosphere and chromosphere through mechanical and Ohmic dissipations. The consequence of this is that the intensity contrast of faculae and network exhibits rather different disc centre-to-limb variation in the two atmospheric regimes \citep{yeo13,yeo19}. The corona, in terms of the structure of the magnetic field and heating processes, differs markedly from the photosphere and chromopshere. We expect the response of the F10.7 to faculae and network to depart from that of TSI more than the other chromospheric indices. Indeed, tests indicate the TSI reconstruction based on the $\fra$ to be poorer at reproducing the PMOD composite than the models based on the $\fca$, $\fla$ and $\fmg$.}

{The proxy approach has its shortcomings, such as the limitations of using chromospheric indices to represent the effects of faculae and network on solar irradiance, as we just discussed \citep[see also][]{yeo14b,yeo19}. While more physical and sophisticated techniques to model TSI exist, such as the 3D magnetohydrodynamics simulation-based model presented by \cite{yeo17b}, there is no obvious way to incorporate the YSK model into such TSI modelling approaches. It is straightforward, however, to incorporate the YSK model into the proxy approach. And as noted earlier, the resulting TSI reconstruction reproduces observed TSI variability reasonably closely, and just as well as existing models based on similar data. This is sufficient for our objective, which is to examine what the observations we made with the YSK model in Sect. 4.2 might imply for solar and stellar brightness variations through a TSI model that incorporates the YSK model.}

For the same reason that a particular level of $S$ does not correspond to a unique value of $F$, discussed in Sect. \ref{analysis2}, $S$ does not uniquely map to TSI. Just as we did in that section to elucidate the overall trend in $F$ with $S$, we also examined how FB, SD, the ratio of the two, and TSI vary with $\sa$ by looking at the solar cycle amplitude.

As with $\Delta{}F$, $\Delta{}\tsi$ represents the deviation in TSI from the 2008 solar cycle minimum level. We define the cycle amplitude of FB, SD, and TSI, {denoted by} $\afb$, $\asd$ and $\atsi$, as the value of the three-year running mean of $k_1\Delta{}\mfsaysk$, $k_2{}\ps$ and $\Delta{}\tsi$ at cycle maxima. In Fig. \ref{fsrtsicyclestrength}, we chart $\afb$, $\left|\asd\right|$, $\ratiofbsd$ and $\atsi$ as a function of $\asa$. The mean of the values from the three TSI models is depicted (dots). The uncertainty is propagated from the $\fitthree$ between $\mfsaysk$ and $F$ (Table \ref{fittable}), and the formal regression error of $k_1$ and $k_2$.

We found $\afb$ to scale quadratically with $\asa$  (Fig. \ref{fsrtsicyclestrength}a) and $\asd$ to scale linearly with $\asa$  (Fig. \ref{fsrtsicyclestrength}b) such that the ratio, $\ratiofbsd$ decreases with $\asa$  (Fig. \ref{fsrtsicyclestrength}c). Consequently, $A(\Delta{}\tsi)$ increases with $\asa$ at a diminishing rate (Fig. \ref{fsrtsicyclestrength}d). This means, similar to what was noted for the {facular-to-sunspot-area} ratio \citep{foukal93,foukal96,foukal98,chapman97,shapiro14} and {the ratio of the variation in facular indices to sunspot area} (Sect. \ref{analysis2}), {the ratio of facular and network brightening to sunspot darkening} decreases as sunspot area increases. As a result, TSI rises more slowly with sunspot area than if the two quantities were linearly related to one another. This is why even though solar cycle 23 is markedly weaker than cycle 22, TSI rose to comparable levels over both cycles \citep{detoma01}. Comparing solar cycle 23 to cycle 22 (marked in Fig. \ref{fsrtsicyclestrength}d), the $\asa$ ratio is 0.70 and the $\atsi$ ratio is 0.89. Due to the non-linear relationship between TSI and sunspot area, while cycle 23 is, in terms of $\sa$, 30\% weaker than cycle 22, the TSI cycle amplitude is only 11\% weaker.

We remind the reader  that {for} Sun-like stars, going above a certain level of activity, they switch from growing brighter with rising activity to becoming dimmer instead, which is interpreted as the transition from a faculae-dominated to a spot-dominated regime \citep{lockwood07,hall09,shapiro14,radick18}. The Sun appears to be just below this transition, opening up the possibility of studying this phenomena by looking at how solar faculae and network relate to sunspots, and extrapolating the apparent relationship to higher activity levels. We attempt exactly that here by projecting the TSI model, which is based on the empirical model of the relationship between facular and sunspot indices, to higher activity levels.

We had examined $\afb$, $\left|\asd\right|$, $\ratiofbsd$ and $\atsi$ as a function of activity as indicated by $\asa$ (Fig. \ref{fsrtsicyclestrength}). In the stellar studies, activity is characterised by the $\logrhk$. Accordingly, we converted $\mfcasaysk$ to the $S$-index scale (not to be confused with sunspot indices, abbreviated as $S$ in this article) using the conversion relationship reported by \cite{egeland17}, and then the result to $\logrhk$ following the procedure of \cite{noyes84}. In this computation, we assumed a solar $(B-V)$ colour index of 0.653 \citep{ramirez12}. In Fig. \ref{fsrstellar}, we plot $\afb$ (red dots), $\left|\asd\right|$ (blue dots) and $\atsi$ (black dots) again, this time as a function of $\logrhk$. The $\logrhk$ level corresponding to each $\afb$, $\left|\asd\right|$, and $\atsi$ value is given by the value of the three-year running mean of $\logrhk$ at that solar cycle maximum. The data points sit just above $\logrhk$ of $-4.9$. The switch in the activity-dependence of the brightness of Sun-like stars is estimated to be between $\logrhk$ of $-4.9$ and $-4.7$ \citep{lockwood07,hall09,shapiro14,radick18}. To probe what happens to facular brightening, sunspot darkening, and TSI here, we fitted a straight line to $\afb$ (red line) and a quadratic polynomial to $\left|\asd\right|$ (blue curve), and extrapolated both fits to $\logrhk$ of $-4.65$. Then, we calculated the TSI level corresponding to the $\afb$ line and the $\asd$ curve, represented {by the black curve}.

While facular brightening scales linearly with $\logrhk$ (red line, Fig. \ref{fsrstellar}), sunspot darkening (blue curve) rises increasingly rapidly with activity. At $\logrhk$ of about $-4.80$, sunspot darkening rises more rapidly with activity than facular brightening, with the consequence that TSI (black curve) goes from increasing with activity to decreasing instead (dashed lines). To aid this discussion, let us abbreviate this transition in TSI activity-dependence as T1. Eventually, the negative correlation between TSI and activity results in TSI dropping below the 2008 minimum level at $\logrhk$ of about $-4.68$ (dotted lines), denoted as T2 ({recall,} $\atsi$ is defined such that null corresponds to the 2008 minimum level). What happens to TSI at T1 and T2 is further illustrated in the following. Supposing $\logrhk$ varies over a solar cycle as pictured in Fig. \ref{fsrhypothetical}a, using the $\atsi$ extrapolation (black curve, Fig. \ref{fsrstellar}), we computed the corresponding TSI (Fig. \ref{fsrhypothetical}b) in the scenario that activity peaks below T1 (red), at T1 (green), between T1 and T2 (blue), at T2 (purple), and above T2 (cyan). When a given cycle is sufficiently strong, such that activity peaks above T1, TSI starts to dip around the cycle maximum, such that instead of varying along with the activity cycle, TSI exhibits a double-peaked form. It is worth emphasising that we based this analysis on the three-year running mean of the various quantities so as to elucidate the overall trend with activity. This discussion does not apply to the variability at shorter timescales.

The observation here that T1 occurs around the range of activity where the activity-dependence of the brightness of Sun-like stars exhibits a similar switch bolsters the interpretation of what we see in Sun-like stars as the transition from a faculae-dominated to a spot-dominated regime. Since the TSI model is based on the empirical model of the relationship between facular and sunspot indices, this result also implies that the latter accurately captures how solar faculae and {network} relate to sunspots. Converted to $\logrhk$, the $\fca$ composite, which extends four solar cycles, has a mean and standard deviation of $-4.90$ and 0.03, respectively. This means T1, at $\logrhk$ of about $-4.80$, is just over three standard deviations away from the mean observed solar level. This underlines how delicately balanced facular brightening and sunspot darkening are on the Sun \citep[c.f.][]{shapiro16}.

Of course, any extrapolation of empirical relationships needs to be interpreted with care. In this case, the TSI model and the underlying model of the relationship between facular and sunspot indices closely replicate a range of observations (see Sect. \ref{analysis1}, Fig. \ref{fsrtsimodel}, and Table \ref{tsitable}), and the turning point of the $\atsi$ extrapolation (i.e. T1: dashed lines, Fig. \ref{fsrstellar}) is close to the data points, where the extrapolation is the most reliable. {As is evident from the figure, the $95\%$ confidence interval of the $\atsi$ extrapolation (shaded region) remains relatively small around T1.} This gives us confidence that the extrapolation of the TSI model is, at least up to T1, {somewhat reasonable}.

The stellar studies are based on the Str\"omgren $b$ and $y$ photometry of Sun-like stars, not the bolometric photometry. However, as there are no direct observations of the Sun in Str\"omgren $b$ and $y$, deriving an empirical model of solar Str\"omgren $b$ and $y$ as we did for TSI is {currently} not possible. This is a limitation of the current model. {Studies have reported linear relationships for converting TSI to solar Str\"omgren $b$ and $y$ {(see \citealt{radick18} and references therein)}, but these are very approximate. Additionaly, since these relationships are linear, repeating the above analysis on the results of having applied them to the PMOD TSI composite would have no qualitative effect on the results.} It would be possible to model Str\"omgren photometry with a proper model of spectral solar irradiance {\citep[e.g.][]{shapiro16}}, but that is outside the scope of the current study. This does not detract however, from the fact that the TSI model indicates the bolometric photometry of the Sun to behave in an analogous manner to the Str\"omgren $b$ and $y$ photometry of Sun-like stars.

\section{Summary}
\label{summary}

How faculae and {network} relate to sunspots is of interest for the implications for solar and stellar brightness variations. In this study, we probed this relationship by looking at how chromospheric indices and the total magnetic flux enclosed in faculae and network, termed faculae indices, compare to sunspot indices. This makes use of the fact that chromospheric emission is strongly enhanced in plage and network features overlaying photospheric faculae and {network}, such that chromospheric indices are, to a greater or lesser degree, modulated with the amount of faculae and network present. Specifically, we compared the 10.7 cm radio flux, Ca II K $1\ \AA{}$ emission index, Lyman $\alpha$ irradiance, Mg II index and total facular and network magnetic flux to the total sunspot area, international sunspot number and group sunspot number.

We presented an empirical model of the relationship between facular and sunspot indices (Sect. \ref{modelysk}). The model is a modification of the \cite{preminger05,preminger06a,preminger06b,preminger07} model, which describes the variation in chromospheric indices from active region evolution. The relationship between a given chromospheric index and sunspot index is described as the convolution of the latter with a suitable finite impulse response (FIR) filter. The modifications introduced here were aimed at more accurately reflecting what \cite{preminger06a} noted of the FIR filters they had derived empirically from observations, and also to take into account the likely non-linear relationship between plage emission and facular magnetic flux, and between the latter and sunspot prevalence. Taking the current model, we reconstructed the facular indices from the sunspot indices. The model not only replicated most of the observed variability (up to $94\%$), but it is also better at doing so than the \cite{preminger05,preminger06a,preminger06b,preminger07} model, as well as that of \cite{tapping11} and \cite{tapping17} (Sect. \ref{analysis1}).

\cite{tapping11} and \cite{tapping17}, along with \cite{svalgaard10} and \cite{livingston12}, found the relationship between the 10.7 cm radio flux and the total sunspot area and international sunspot number to have changed since solar cycle 23. {The cited studies, in their analyses, had implicitly assumed that the 10.7 cm radio flux at a given time is a function of the sunspot area/number at that time alone (see Sects. \ref{introduction} and \ref{modeltvm}). The magnetic flux in active regions persists, manifest as faculae and {network}, for some time after the sunspots they bear have decayed. As a consequence, the amount of faculae and {network} present at a given time, and therefore the response of the F10.7 to these magnetic structures, is indicated not just by prevailing sunspots, but also sunspots that had emerged in the recent past.} Contrary to these studies, we found no clear evidence of any secular variation in the relationship between the facular and sunspot indices examined over the past four decades (Sect. \ref{analysis1}). The present analysis made use of a model that takes {what we just noted about} active region evolution into account and demonstrated to be more physical than the \cite{tapping11} and \cite{tapping17} model, and is extended to more facular and sunspot indices. Taking this into consideration, the conflicting results from the earlier studies is likely an artefact of data uncertainty and limitations in their analyses and models.

Various studies have noted that the facular-to-sunspot-area ratio diminishes with sunspot area \citep{foukal93,foukal96,foukal98,chapman97,shapiro14}. Comparing the reconstruction of the facular indices from the sunspot indices to the latter, we found the ratio of the variation in chromospheric emission and total facular and network magnetic flux to sunspot area to exhibit the same behaviour, decreasing with increasing sunspot area (Sect. \ref{analysis2}).

Making use of the fact that chromospheric indices are reasonable proxies of the effect of faculae and {network} on solar irradiance, we examined the implications for solar and stellar brightness variations by means of an empirical model of total solar irradiance (TSI). The TSI model, which incorporates our model of the relationship between chromospheric indices and total sunspot area, indicates that the ratio of the TSI excess from faculae and network and the deficit from sunspots also decreases with increasing sunspot area. The consequence is that TSI rises with sunspot area more slowly than if the two quantities are linearly related to one another (Sect. \ref{analysis3}). TSI rising with activity at a diminishing rate explains why TSI rose to comparable levels over solar cycles 22 and 23, even though cycle 23 is significantly weaker than cycle 22. The current study extended and improved upon what was noted earlier of the {facular-to-sunspot-area} ratio through an examination of independent data sets over multiple cycles (as noted in the introduction, a proper study of the relationship between facular and sunspot area over multiple solar cycles is still lacking).

We extrapolated the trend in facular and network brightening, sunspot darkening, and TSI with activity to higher activity levels, over the range where Sun-like stars are observed to switch from growing brighter with rising activity to becoming dimmer instead. The projection indicates that in this activity range, sunspot darkening will gradually rise faster with activity than faculae and network brightening, such that the activity-dependence of TSI exhibits a similar switch. This bolsters the interpretation of the dichotomy in the relationship between the brightness and activity of Sun-like stars as the transition from a faculae-dominated to a spot-dominated regime. This result also suggests that our model of the relationship between facular and sunspot indices, which underlies the TSI model, accurately captures how solar faculae and network relate to sunspots.

\begin{acknowledgements}
We made use of the \cite{balmaceda09} total sunspot area and PSI composites (www2.mps.mpg.de/projects/sun-climate/data.html), the international sunspot number (www.sidc.be/silso), the group sunspot number (original; www.sidc.be/silso, revision by \cite{chatzistergos17}; www2.mps.mpg.de/projects/sun-climate/data.html), the Penticton 10.7 cm radio flux record (lasp.colorado.edu/lisird), the \cite{bertello16} Ca II K $1\ \AA$ emission index composite (solis.nso.edu/0/iss/), the \cite{machol19} Lyman $\alpha$ irradiance composite (lasp.colorado.edu/lisird), the IUP Mg II index composite (www.iup.uni-bremen.de/gome/gomemgii.html) and the PMOD TSI composite (ftp://ftp.pmodwrc.ch/pub/data/). The reconstruction of the 10.7 cm radio flux presented in this study is available for download at www2.mps.mpg.de/projects/sun-climate/data.html. We received funding from the German Federal Ministry of Education and Research (project 01LG1209A), the European Research Council through the Horizon 2020 research and innovation programme of the European Union (grant agreement 695075) and the Ministry of Education of Korea through the BK21 plus program of the National Research Foundation.
\end{acknowledgements}

\bibliographystyle{aa}
\bibliography{references}

\end{document}